\documentclass{epl}

\title{Comment on ``Statistical mechanics of CDMA multiuser demodulation"}
\author{Hidetoshi Nishimori}
\institute{
 Department of Physics, Tokyo Institute of Technology,
 Oh-okayama, Meguro-ku, Tokyo 152-8551, Japan
}
\pacs{84.40.Ua}{Telecommunications: signal transmission and processing;
communication satellites}
\pacs{05.50.+q}{Lattice theory and statistics}
\pacs{89.70.+c}{Information science}

\begin{document}

\maketitle

In a recent paper~\cite{Tanaka}, Tanaka formulated and solved
a model of CDMA multiuser demodulation by applying the
theory of spin glasses, the replica method in particular.
It is shown in the present comment that some of his results can be
derived without recourse to the replica method;
conclusions from the somewhat tricky replica method are
justified rigorously.
It also becomes clear that there is no finite-size effect
in the internal energy for the MPM demodulator and that the
structure of the phase space is simple; the latter goes beyond the
local stability argument of the AT line analysis.

The first of our result is a generalization of the identity $m=q$
for the MPM demodulator derived by Tanaka using the replica method:
The functional identity $P_m(x)=P_q(x)$ will be proved for MPM.
Here $P_m(x)~(-1\le x \le 1)$ is the distribution function of the order
parameter corresponding to magnetization and $P_q(x)$ denotes
the distribution of the spin glass order parameter.

The proof follows the same line as in the spin glass case
\cite{NS,Nishimori2001}.
The function $P_m(x)$ is defined as
 \begin{eqnarray}
 P_m(x)&=&
    \frac{1}{2^{Np+N}}\left( \frac{\beta_{\rm s}}{2\pi}\right)^{p/2}
  \sum_{\eta} \int \prod_t dr^t\, \sum_{\xi}
  \exp \left\{-\frac{\beta_{\rm s}}{2}
   \sum_t (r^t -\frac{1}{\sqrt{N}}\sum_i \eta_i^t \xi_i)^2 \right\}
    \nonumber\\
   &\cdot &
   \frac{\sum_{s} \delta (x-\frac{1}{N}\sum_i \xi_i S_i)
  \exp \left\{-\frac{\beta}{2}
   \sum_t (r^t -\frac{1}{\sqrt{N}}\sum_i \eta_i^t S_i)^2 \right\}}
   {\sum_{s} 
  \exp \left\{-\frac{\beta}{2}
   \sum_t (r^t -\frac{1}{\sqrt{N}}\sum_i \eta_i^t S_i)^2 \right\}}.
    \label{Pm}
 \end{eqnarray}
Here $r^t$ is the received signal scaled by $\sqrt{N}$,
$r^t=y^t/\sqrt{N}$,
$\eta_i^t$ is the spreading code sequence, $\xi_i$ is the
original information symbol of user $i$, and $S_i$ is
the dynamical variable for demodulation.
Note that the order parameter $m$ in Tanaka's article stands for
the average overlap of the original symbol $\xi_i$ and the thermal
expectation value of the demodulation variable $S_i$.
Then the function $P_m(x)$ represents the distribution of his
order parameter.
The function $P_m(x)$ is equivalent to the usual magnetization
appearing in the theory of spin glasses:
One can confirm it by the gauge transformation $\eta_i^t\to
\eta_i^t \xi_i$ and $S_i\to S_i \xi_i$ for given $\xi_i$.
Then the expression after $\sum_{\xi}$ in eq.~(\ref{Pm}) becomes
independent of $\xi_i$, so that this sum over $\xi$ can be eliminated.
The result is that $P_m(x)$ measures the magnetization distribution
of the system with dynamical variables $S_i$.

The distribution function $P_q(x)$ is defined similarly
with the second line of eq.~(\ref{Pm}) replaced by
 \begin{equation}
   \frac{\sum_{s,\sigma} \delta (x-\frac{1}{N}\sum_i S_i\sigma_i)
  \exp \left\{-\frac{\beta}{2}
   \sum_t (r^t -\frac{1}{\sqrt{N}}\sum_i \eta_i^t S_i)^2 
  -\frac{\beta}{2}
   \sum_t (r^t -\frac{1}{\sqrt{N}}\sum_i \eta_i^t \sigma_i)^2
   \right\}}
   {\sum_{s,\sigma} 
  \exp \left\{-\frac{\beta}{2}
   \sum_t (r^t -\frac{1}{\sqrt{N}}\sum_i \eta_i^t S_i)^2 
  -\frac{\beta}{2}
   \sum_t (r^t -\frac{1}{\sqrt{N}}\sum_i \eta_i^t \sigma_i)^2
   \right\}}.
    \label{Pq}
 \end{equation}
Comparison of eqs.~(\ref{Pm}) and (\ref{Pq}) immediately reveals
the relation $P_m(x)=P_q(x)$ under the MPM condition
$\beta =\beta_{\rm s}$ since one of the sums in the denominator
of eq.~(\ref{Pq}) cancels with the sum over $\xi$ in the
first line of eq.~(\ref{Pm}).

It is well established that the distribution of magnetization
is simple, that is, $P_m(x)$ is composed of at most two delta
functions, and therefore the function $P_q(x)$ should
also be simple.
This implies that the structure of the phase space is simple
and there is no replica symmetry breaking for MPM demodulator,
in agreement with the AT stability analysis in replica
calculations.
Indeed the simple structure of $P_q(x)$ means more:
The AT analysis is for local stability against RSB perturbations
whereas the simple $P_q(x)$ suggests that the whole phase
space is simple, a result on the global structure.
The identity $m=q$ derived by Tanaka can be proved
by integration of $xP_m(x)$ and  $xP_q(x)$ over $0\le x\le 1$.

Along the same line of reasoning, it is possible to evaluate the
internal energy $U$ for MPM demodulator without using replicas.
The definition of the internal energy is almost the same as
eq.~(\ref{Pm}), the only difference being that the delta function
in the numerator of the second line is replaced with 
partial derivative $\partial /\partial\beta$.
It then follows that the sums over $\xi$ and $S$ cancel under the
MPM condition $\beta=\beta_{\rm s}$.
It is straightforward to evaluate the resulting expression by
carrying out the integral over $r^t$ first and then
taking derivative by $\beta$.
The result is
 \begin{equation}
    U =-\frac{p}{2\beta}.
    \label{energy}
 \end{equation}
This agrees with the replica calculation as can be verified by taking
the $\beta$-derivative of the free energy, Tanaka's eq.~(10).
It is remarkable that our eq.~(\ref{energy}) is applicable to any
finite-size system whereas Tanaka's energy is for the thermodynamic
limit.
Agreement of the two results means that finite-size effects are
completely absent in the MPM internal energy
averaged over quenched randomness.

We have shown that some of the important results of Tanaka can
be derived without replicas as long as the MPM demodulator is concerned.
Application of the present method to the MAP case is an interesting
but difficult future problem.

\acknowledgments
This work was supported partly by Sumitomo Foundation.

\end{document}